\documentstyle{article}
\begin{document}
\title{Neutral Particles and Super Schwinger Terms}
\author{{\bf C.Ekstrand}\\Department of Theoretical Physics, \\Royal Institute of
Technology, \\S-100 44 Stockholm, Sweden}
\date{}
\maketitle
\newcounter{proposition}
\setcounter{proposition}{1}

\newcounter{theorem}
\setcounter{theorem}{1}
\newcounter{lemma}
\setcounter{lemma}{1}
\renewcommand{\thetheorem}{\arabic{theorem}}
\renewcommand{\thelemma}{\arabic{lemma}}
\newenvironment{proof}{\newline {\bf Proof } }{ \newline \newline }
\newenvironment{proposition}{\newline \newline {\bf Proposition }{\bf\theproposition } \em }{\newline \addtocounter{proposition}{1}}

\newenvironment{theorem}{\newline \newline {\bf Theorem }{\bf\thetheorem } \em }{ \newline\addtocounter{theorem}{1}}

\newenvironment{lemma}{\newline \newline {\bf Lemma }{\bf \thelemma }\em }{\newline \addtocounter{lemma}{1}}

\newcommand{\eq}{\begin{equation}}
\newcommand{\eqend}{\end{equation}}
\newcommand{\eqa}{\begin{eqnarray}}
\newcommand{\eqaend}{\end{eqnarray}}
\newcommand{\nonu}{\nonumber \\ \nopagebreak}
\newcommand{\Ref}[1]{(\ref{#1})}

\newcommand{\B}{{\cal B}}
\newcommand{\F}{{\cal F}}
\newcommand{\cL}{{\cal L}}
\newcommand{\U}{{\cal U}}
\begin{abstract}
 ${\bf{Z}}_2$-graded Schwinger terms for neutral particles in 1 and 3 space dimensions are considered. 
\end{abstract}
\section{Introduction}
 In this paper we consider the canonical commutation relations (CCR) algebra and the canonical anticommutation relations (CAR) algebra. To some extent it is possible to treat the two algebras in parallel.
 This has been done in several places. Close to our spirit is \cite{GL2}.
 In this reference the CCR 
and the CAR algebra are generalized to a superalgebra -- the canonical 
supercommutation relations (CSR) algebra, which describes charged particles in a ${\bf{Z}}_2$-graded fashion.

 Related to these algebras is the concept of Schwinger terms 
\cite{GI,SW}. It is cocycles that arises when trying to implement Bogoliubov 
transformations in a physically important representation. They have been 
obtained for the CCR algebra in \cite{LA}, for the CAR algebra in \cite{L} and for the CSR algebra in \cite{GL2}, for example. Schwinger terms for neutral bosons and fermions have been is calculated in \cite{AR}.

What lacks is a calculation of a ${\bf{Z}}_2$-graded Schwinger term for neutral particles. In 
section \Ref{sec:IAS} we perform such a calculation on an abstract basis by 
deriving (in section \Ref{sec:SSR}) a ${\bf{Z}}_2$-graded generalization of 
the fact \cite{A2} the 
CAR algebra is $\ast$-isomorphic with the self-dual canonical anticommutation 
relations (SDC) algebra, used to describe neutral particles. Finite 
Bogoliubov transformations are considered in section \Ref{sec:FBT}. In section \Ref{sec:APR}, a physical realization is made which allows concrete computations of ${\bf{Z}}_2$-graded Schwinger term for neutral particles in odd dimensional space. Explicit results are given for 1 and 3 space dimensions. Finally, in 
section \Ref{sec:IGN}, the formalism is extended to include Grassmann numbers. 

\section{The SSR and CSR algebra}
\label{sec:SSR}
 Assume that the boson space $h_{\bar{0}}$ and the fermion space $h_{\bar{1}}$ are complex, infinite dimensional, separable Hilbert spaces. We will in general be interested in the ${\bf{Z}}_2$-graded space $h=h_{\bar{0}}\oplus h_{\bar{1}}$. Its scalar product will be denoted by $( \cdot ,\cdot )$. Let $J$ be a conjugation in $h$ (an antilinear norm-preserving operator whose square is the identity) that is even (i.e. it is commuting with the Klein operator $\gamma= P_{\bar{0}}-P_{\bar{1}}$, where $\gamma = P_{\bar{0}}-P_{\bar{1}}$ and $P_{\alpha}$, $\alpha=\bar{0},\bar{1}$, denotes the orthogonal projection onto $h_{\alpha}$). Let $P$ be an even orthogonal projection operator that satisfies $JP = (1-P)J$. The physical interpretation we have in mind is that particles are described by vectors in $Ph$ and anti-particles by vectors in $(1-P)h$. Since the algebraic relations for creation and annihilation operators differ between bosons and fermions, it is useful to introduce the operator $q=P-\gamma (1-P)$. In terms of it, these relations (considered in \cite{AR}) can be extended to 
\eqa
\label{eq:SSR}
B^{\dagger }(c_1f_1+c_2f_2) & = & c_1B^{\dagger }(f_1)+c_2B^{\dagger }(f_2) 
\nonu
B^{\dagger }(f) & = & B(qf)^\ast \nonu
\left[ B(f_1),B^{\dagger }(f_2)\right]_s & = & \left( f_1,f_2\right) {\bf 1}
\nonu
B(f) & = & B^{\dagger }(qJ f)
\eqaend
with 
\eq
\mbox{\small{deg}}(B^{\dagger }(f))=\mbox{\small{deg}}(f),
\eqend
where the supercommutator has been used (basic definitions concerning super algebras can be found in the Appendix). The asterix  denotes the adjoint operation in $h$. Replacing the last equation in \Ref{eq:SSR} with the fact that the commutator of two $B$ operators vanish gives the well known relations for the generators of the CSR algebra (which will be defined below for the \lq half\rq -space $Ph$). Our choice of the last equation in \Ref{eq:SSR} relates the creation of a particle (anti-particle) with the annihilation of a corresponding anti-particle (particle). This means that \Ref{eq:SSR} describes neutral particles, see \cite{RU1}. We define the self-dual canonical super commutator relations (SSR) algebra ${\cal A}_{SSR}$ over $h$ to be the complex $\ast$-algebra generated by an identity $\bf{1}$ and the symbols $B^{\dagger }(f),B(f)$; $f\in h$, obeying the relations generated by \Ref{eq:SSR}. The reason for referring to $B^{\dagger }(f)$ and $B(f)$ as symbols is that we have not yet given any representation of ${\cal A}_{SSR}$ in terms of which they can be interpreted as creation and annihilation operators. To find such a representation, we set $\tilde{h}=Ph$ (we could equally well have defined $\tilde{h}$ to be $(1-P)h$ since any two complex infinite dimensional separable Hilbert spaces are isomorphic) and define the canonical super commutation relations (CSR) algebra ${\cal A}_{CSR}$ over $\tilde{h}$ to be the complex $\ast$-algebra generated by an identity $\bf{1}$ and the symbols $a^{\dagger }(\tilde{f}),a(\tilde{f})$; $\tilde{f}\in \tilde{h}$, obeying the relations generated by
\eqa
\label{eq:CSR}
a^{\dagger }(c_1\tilde{f}_1+c_2\tilde{f}_2) & = & 
c_1a^{\dagger }(\tilde{f}_1)+c_2a^{\dagger }(\tilde{f}_2) 
\nonu
a^{\dagger }(\tilde{f}) & = & a(\tilde{f})^\ast \nonu
\left[ a(\tilde{f}_1),a^{\dagger }(\tilde{f}_2)\right]_s & = & 
\left( \tilde{f}_1,\tilde{f}_2\right) {\bf 1}
\nonu
\left[ a(\tilde{f}_1),a(\tilde{f}_2)\right]_s & = & 0
\eqaend
with
\eq
\mbox{\small{deg}}(a^{\dagger }(\tilde{f}))=\mbox{\small{deg}}(\tilde{f}).
\eqend
 This is recognized as the algebra that describes charged bosons and fermions when considered as vectors in $\tilde{h}$. Since it has been considered earlier in the literature \cite{GL2} we can find useful information about ${\cal A}_{SSR}$ if it can be related to ${\cal A}_{CSR}$. In fact, the following lemma holds:
\begin{lemma}
${\cal A}_{SSR}$ and ${\cal A}_{CSR}$ are $\ast$-isomorphic
\end{lemma}
\begin{proof}
Define a mapping $\chi$ by 
\eq
\label{eq:SEC}
\chi \left(B^{\dagger }\left(f\right)\right)  = a^{\dagger }(Pf)
-a(\gamma J(1-P) f)
\eqend
It is then straight forward to check that $\chi$ can be extended to a 
$\ast$-isomorphism. 
\end{proof}
 We will now consider a representation of the algebra in  \Ref{eq:CSR} 
called the Fock-Cook representation \cite{GL2}. 
The representation space is the Fock space $\F _{\gamma }(\tilde{h}) = 
\F _B(\tilde{h}_{\bar{0}})\otimes \F _F(\tilde{h}_{\bar{1}})$, $\tilde{h}_{{\alpha }}:=P_{\alpha }\tilde{h}$, with the naturally induced ${\bf{Z}}_2$-structure. In terms of the bosonic and fermionic creation and annihilation 
operators (in the Fock-Cook representation), the representation is given by:
\eq
\label{eq:FRE}
\hat{a}^{\dagger }(\tilde{f})\equiv
\pi \left(a^{\dagger }(\tilde{f})\right) =  
\hat{a}^{\dagger }_B ( P_{\bar{0}}\tilde{f})\otimes {\bf 1}_{\bar{1}} 
+{\bf 1}_{\bar{0}} \otimes \hat{a}^{\dagger }_F ( P_{\bar{1}}\tilde{f}) 
\eqend 
and $\hat{a}(\tilde{f})=\hat{a}^{\dagger }(\tilde{f})^{\ast}$. 
The vacuum $ |\Omega \!\! >= |\Omega _B\!\! >\otimes\,\,\,|\Omega _F\!\! >$ is 
the vector characterized 
 by $\hat{a}(\tilde{f})|\Omega \!\! > =0$, $\forall \tilde{f}\in\tilde{h}$.
 Through lemma 1 this construction provides us with a (physically interesting) representation of 
 \Ref{eq:SSR} according to 
\eqa
\label{eq:SE0}
\hat{B}^{\dagger }(f) & = & \hat{a}^{\dagger }(Pf)-
\hat{a}(\gamma J(1-P)f)
\nonu
\hat{B}(f)            & = & \hat{a}(Pf)+
\hat{a}^{\dagger }(J(1-P)f) ,
\eqaend
 see \cite{AR} for the corresponding ungraded representation.

\section{Implementation and Schwinger terms}
\label{sec:IAS}
 The implementation problem for the representation in \Ref{eq:SE0} will now be considered. We will prove a corresponding theorem to the one proved by Shale \cite{SH} (for charged bosons) and by Shale and Stinespring \cite{SS} (for charged fermions). However, the language of Ruijsenaars \cite{RU1} will be used.

 Consider a bounded operator $\tilde{X}$ on $\tilde{h}$ having the form 
\eq
\tilde{X}=\sum_{n=1}^\infty \tilde{\lambda } _n\tilde{f}_n(\tilde{g}_n, \cdot )
\eqend
with $\{ \tilde{f}_n\} _{n=1}^\infty$, $\{ \tilde{g}_n\} _{n=1}^\infty$ 
orthonormal systems in $\tilde{h}$ and $\{ \tilde{\lambda } _n\} _{n=1}^\infty$ a set of complex numbers.
The operator 
\eq
 \label{eq:AA1}
\tilde{X}\hat{a}^{\dagger }\hat{a} = 
\sum_{n=1}^\infty \tilde{\lambda } _n\hat{a}^{\dagger }(\tilde{f}_n) 
\hat{a}(\tilde{g}_n) 
\eqend 
is well defined on the dense subspace ${\cal D}(\tilde{h})\subset 
\F _{\gamma }(\tilde{h})$ defined as the linear span of elements 
\eq
\hat{a}^{\dagger }(\tilde{f}_1)\hat{a}^{\dagger }(\tilde{f}_2)...
\hat{a}^{\dagger }(\tilde{f}_{N_0})|\Omega \!\! > \quad N_0\in {\bf N}, 
\tilde{f_n}\in \tilde{h}
\eqend
 where the multiplication of the creation operators is either the symmetric 
or antisymmetric exterior product depending on the degree of the vectors 
under consideration.
 For a given conjugation $\tilde{J}$ on $\tilde{h}$ we define 
\eqa
\label{eq:AA2}
 \tilde{X}\hat{a}^{\dagger }\hat{a}^{\dagger } & = & 
\sum_{n=1}^\infty \tilde{\lambda} _n\hat{a}^{\dagger }(\tilde{f}_n) 
\hat{a}^{\dagger }(\tilde{J}\tilde{g}_n)
\nonu
\tilde{X}\hat{a}\hat{a} & = & 
\sum_{n=1}^\infty \tilde{\lambda }_n\hat{a}(\tilde{J}\tilde{f}_n) 
\hat{a}(\tilde{g}_n). 
\eqaend 
 The following lemma was proven in \cite{GL2}: 
\begin{lemma}
Let $\tilde{X}$ be a bounded  
operator on $\tilde{h}$. The necessary and sufficient condition for 
the operators $\tilde{X}\hat{a}\hat{a}$ and 
$\tilde{X}\hat{a}^{\dagger }\hat{a}^{\dagger }$ to be well defined  on  
${\cal D}(\tilde{h})$ is that $\tilde{X}$ is a Hilbert-Schmidt operator.
\end{lemma}

 Consider then bounded operators $X$ on $h$ of form 
\eq
X=\sum_{n=1}^\infty \lambda _nf_n(g_n, \cdot ) ,
\eqend
where $\{ f_n\} _{n=1}^\infty$ and $\{ g_n\} _{n=1}^\infty$ are 
orthonormal systems in $h$ and $\{ \lambda _n\} _{n=1}^\infty$ is a set of complex numbers. For a moment, we will restrict to the case when 
$X$ is an operator such 
that 
\eq
\label{eq:NOD}
\frac{1}{2}\sum_{n=1}^\infty \lambda _n\hat{B}^{\dagger }
(f_n)\hat{B}(g_n)
\eqend
exists as an operator on ${\cal D}(\tilde{h})$ and has finite vacuum 
expectation value.
 This is equivalent with the fact that $X$ is a trace class operator. The operator $\hat{d\Gamma } (\cdot )$ is defined by
\eq
\label{eq:DG}
\hat{d\Gamma } (X)=\,\,\, :\frac{1}{2}\sum_{n=1}^\infty \lambda _n\hat{B}^{\dagger }
(f_n)\hat{B}(g_n): ,
\eqend
where $:\quad :$ denotes normal ordering defined as 
the argument subtracted by its vacuum expectation value.
 It follows that
\eq
\label{eq:GX}
\hat{d\Gamma } (X)=\hat{d\Gamma }(qJ \epsilon (X)^{\ast}qJ), 
\eqend
where
\eq
\epsilon (\sum_{n=1}^\infty \lambda _nf_n(g_n, \cdot ))=
\sum_{n=1}^\infty (-1)^{<f_n,g_n>}\lambda _nf_n(g_n, \cdot ) 
\eqend
and $<f_n,g_n>:=\mbox{\small{deg}}(f_n)\cdot\mbox{\small{deg}}(g_n)$ (see Appendix). It is important to note that the last relation in \Ref{eq:SSR} was necessary for the derivation of \Ref{eq:GX}. Thus, no corresponding relation exist for charged particles. From now on, we will only consider operators $X$ fulfilling the so-called self-duality condition \cite{EL}:
\eq
\label{eq:SDC}
X=qJ \epsilon (X)^{\ast}qJ .
\eqend 
 Inserting  \Ref{eq:SE0} in \Ref{eq:DG} gives then:
\eq
\label{eq:DGA}
\hat{d\Gamma }(X)=(PXP)\hat{a}^{\dagger }\hat{a} 
+ \frac{1}{2}(PXJ\tilde{J}P) \hat{a}^{\dagger }\hat{a}^{\dagger }
- \frac{1}{2}(\gamma P\tilde{J}JXP)\hat{a}\hat{a}, 
\eqend
where the relation 
\eq
\frac{1}{2}\sum_{n=1}^\infty \lambda _n\hat{a}^{\dagger }(Pf_n) \hat{a}(Pg_n)=
\,\,\, :\frac{1}{2}\sum_{n=1}^\infty \lambda _n\hat{a}(\gamma PJ f_n) 
\hat{a}^{\dagger }(PJg_n):
\eqend
 has been used. With this alternative form it is possible to extend the domain of $\hat{d\Gamma } (\cdot )$. For this, we introduce the semi-group $g_2$ of bounded self-dual operators $X$ such that $PX(1-P)$ and $(1-P)XP$ are Hilbert-Schmidt.
\begin{proposition}
Let $X$ be a bounded self-dual operator on $h$. 
The necessary and sufficient condition for $\hat{d\Gamma }(X)$ in  
\Ref{eq:DGA} to be well defined on ${\cal D}(\tilde{h})$ is that $X\in g_2$. 
 Further, for $X,Y\in g_2$, 
\eq
\label{eq:DGB}
\left[ \hat{d\Gamma }(X),\hat{B}^{\dagger }(f)\right] _s  =  
\hat{B}^{\dagger }(Xf) 
\eqend
\eq 
\label{eq:DGG}
\left[ \hat{d\Gamma }(X),\hat{d\Gamma }(Y)\right] _s = \hat{d\Gamma }(
\left[ X,Y\right] _s)-\frac{1}{2}c(X,Y){\bf 1},
\eqend
with Schwinger term 
\eq
\label{eq:ST}
c(X,Y)= \mbox{\rm Str} \left( (1-P)XPY -(-1)^{<X,Y>}(1-P)YPX 
\right)
\eqend
where $\mbox{\rm Str}(\cdot )=\mbox{\rm Tr}(\gamma\cdot )$ and $\mbox{\rm Tr}$ is the Hilbert space trace on $h$.
\end{proposition}
\begin{proof}
 $\hat{d\Gamma }(X)$ is well defined if and only if all operators on the right hand side are well defined. For the first term this is clear. According to lemma 1, the necessary and sufficient condition for the last two to be well defined is that $PXJ\tilde{J}P$ and $\gamma P\tilde{J}JXP$ are Hilbert-Schmidt. The first part of the proposition follows now from the fact that the algebra of Hilbert-Schmidt operators is an ideal in the set of bounded operators, see for example \cite{S}. By use of eq. \Ref{eq:CSR}, \Ref{eq:SE0} and \Ref{eq:DGA}, it is straight forward to verify eq. \Ref{eq:DGB}. The calculations are simple but long and will therefore be omitted. Similar, the relation in \Ref{eq:DGG} can be proved by use of eq. \Ref{eq:CSR} and \Ref{eq:DGA}. The Schwinger term can be obtained by observing that the first term on the right hand side of eq. \Ref{eq:DGG} has vanishing vacuum expectation value. It implies:
\eq
c(X,Y)=-2<\!\!\Omega |
\left[ \hat{d\Gamma }(X),\hat{d\Gamma }(Y)\right] _s 
|\Omega \!\! >,
\eqend
from which eq. \Ref{eq:ST} can be proven. Again, the long, but simple computations will be omitted.
\end{proof}
 Notice that the Schwinger term that arises in eq. \Ref{eq:ST} is precisely one half of the corresponding term for charged particles. Another important distingtion from the charged particle case is that proposition 1 is only true for operators satisfying the self-duality condition. In fact, eq. \Ref{eq:GX} implies that 
\eq
\hat{d\Gamma }(X)=\frac{1}{2}\hat{d\Gamma }\left( X+qJ \epsilon (X)^{\ast}qJ\right)
\eqend
leading to
\eq
\left[ \hat{d\Gamma }(X),\hat{B}^{\dagger }(f)\right] _s  =  
\hat{B}^{\dagger }\left( \frac{1}{2}\left( X+qJ \epsilon (X)^{\ast}qJ\right)f\right) 
\eqend
which is not equal to $\hat{B}^{\dagger }(Xf)$ unless $X$ is self-dual. A 
similar argument can be made regarding eq. \Ref{eq:DGG} and \Ref{eq:ST}.

 By representing the neutral creation and annihilation operators in terms 
of the corresponding operators for charged particles, eq. \Ref{eq:SE0}, 
the question of the extension to unbounded operators of the results in this section is easy to answer. In fact, it is in complete analogy with the corresponding extension for charged particles, considered in \cite{GL2}. The results in the next section and in section \Ref{sec:IGN} can be extended to unbounded operators in the same way.

\section{Bogoliubov transformations}
\label{sec:FBT}
 The necessary and sufficient condition for an even operator $U$ on $h$ 
to induce an automorphism $\alpha _U$ of ${\cal A}_{SSR}$ such that 
\eq
\alpha _U(B^{\dagger }(f))=B^{\dagger }(Uf)
\eqend
is that $\alpha _U$ preserves the relations in \Ref{eq:SSR}, or equivalently, 
that $U$ commutes with $J$ and fulfills $qUqU^{\ast }=1$. We refer to the 
induced transformation as a Bogoliubov transformation. Further, $\alpha _U$ 
is said to be unitary implementable (in the representation under 
consideration) if there exists a unitary operator $\hat{\Gamma } (U)$ such 
that 
\eq  
\hat{\Gamma } (U)B^{\dagger }(f)\hat{\Gamma } (U)^{\ast }=B^{\dagger }(Uf)
\eqend
It is known \cite{AR} that this is equivalent with the condition that $(1-P)UP$ 
and $PU(1-P)$ should be Hilbert-Schmidt operators. The following proposition 
was proved for the charged particle case in ref. \cite{GL2}.
\begin{proposition}
Let $X\in g_2$ be an even operator satisfying $X^{\ast }=qXq$. Then 
\eq
e^{it\hat{d\Gamma }(X)}B^{\dagger }(f)e^{-it\hat{d\Gamma }(X)} = 
B^{\dagger }(e^{itX}f) 
\eqend
\eq
\label{eq:EGE}
e^{it\hat{d\Gamma }(X)}\hat{d\Gamma }(Y)e^{it\hat{d\Gamma }(X)} = 
\hat{d\Gamma }(e^{itX}Ye^{-itX})-\frac{1}{2}b(tX,Y)
\eqend
holds on ${\cal D}(\tilde{h})$ for $t\in {\bf R},f\in h,Y\in g_2$, where
\eq
b(X,Y)=i\int _0^1ds\,\, c(X,e^{isX}Ye^{-isX}).
\eqend
Further, if $Y\in g_2$ is even and fulfills $Y^{\ast }=qYq$ then the following relation holds on ${\cal D}(\tilde{h})$:
\eq 
e^{it\hat{d\Gamma }(X)}e^{it\hat{d\Gamma }(Y)}e^{-it\hat{d\Gamma }(X)}=
e^{-i\frac{1}{2}b(X,Y)}e^{it\hat{d\Gamma }(^{itX}Ye^{-itX})}.
\eqend
\end{proposition}
\begin{proof}
Using the operator relation 
\eq
e^{iA}Be^{-iA}=\sum _{n=0}^{\infty }\frac{i^n}{n!}\underbrace{[A,[A,...[A,}_{\mbox{\footnotesize{$n$ times}}}B]...]] 
\eqend
together with \Ref{eq:DGB}, \Ref{eq:DGG} and \Ref{eq:ST}, the first three equations are 
easily seen to be true. The last relation is obtained by exponentiation of 
\Ref{eq:EGE} for $t=1$. 
\end{proof}
From the relation 
\eq
\hat{d\Gamma }(X)^{\ast }=\hat{d\Gamma }(qX^{\ast }q),\quad X\in g_2
\eqend
it is seen that if $X$ fulfills $X^{\ast }=qXq$ then $\hat{d\Gamma }(X)$ 
is self-adjoint and $e^{i\hat{d\Gamma }(X)}$ is unitary. Thus, for 
$X$ as in the proposition, the operator $U=e^{itX}$ defines a unitary 
implementable Bogoliubov transformation $\alpha _U$.
 
\section{A Physical Realization}
\label{sec:APR}
 We will now consider the case when $h=L^2({\bf R}^d)\otimes V$, where the finite dimensional ${\bf Z}_2$-graded space $V$ induces a natural grading in $h$ by: $h_{\alpha}=L^2({\bf R}^d)\otimes V_{\alpha}$. The Schatten ideal $\B _{2p}$ is defined as the set of bounded operators $X$ such that $\mbox{\rm Tr}(X^{\ast }X)^p$ is convergent, see \cite{S} for instance. Especially, $\B _1$ and $\B _2$ are the trace class and Hilbert-Schmidt operators, respectively. The trace of a pseudo-differential operator (PSDO) that is trace class can be calculated as:
\eq
\mbox{\rm Tr}(X)=\frac{1}{(2\pi)^d} \int d^dxd^dq \,\,\mbox{tr}_V \sigma (X)(q,x),
\eqend  
where $\mbox{tr}_V$ is the trace on $V$ and $\sigma (X)$ denotes the symbol of $X$ as a PSDO, see \cite{T} for the basic properties of PSDO's. From this formula it follows that a PSDO with compact support in configuration space is in $\B _{2p}$ if and only if its symbol is of order less than $-d/2p$. 

 Assume that the decomposition $h=Ph\oplus (1-P)h$ is determined by the sign $\epsilon $ of the hamiltonian so that $\epsilon =1$ on $Ph$ and $\epsilon =-1$ on $(1-P)h$. On the space of eigenvectors belonging to the zero eigenvalue, we define $\epsilon $ to be 1. The condition that a bounded self-dual operator $X$ is in $g_2$ can then be written as $[\epsilon ,X]\in\B _2$. We will from now on only consider the case when $X$ represents an infinitesimal gauge transformation. We therefore assume that it is a multiplication operator: $Xf(x)=X(x)f(x)$, that has compact support in the configuration space and  commutes with the gamma matrices. This implies that $X$ and $\epsilon $ are PSDO of order 0 that commutes to highest order. Thus, their commutator is a PSDO of order $-1$ and $[\epsilon ,X]\in\B _{d+1}$. Therefore, $X\in g_2$ is only true for $d=1$. This corresponds to the fact that the current $\hat{d\Gamma }(X)$ is only well defined in 1 space dimension. In higher dimensions, an additional renormalization, apart from the normal ordering, has to be performed. This is in complete analogy with the case for charged particles. 

 We will now also take into account that the implementer of an infinitesimal gauge transformation is in fact not given by the current, but rather by the Gauss law commutators
\eq
G(x,\epsilon )=\hat{d\Gamma }(X)+\cL _X,
\eqend
 if we restrict to the consider $X\in g_2$. The Lie derivative $\cL _X$ 
acts on functionals  $m(\epsilon )$ as 
\eq
(\cL _Xm)(\epsilon )= \frac{d}{dt}\left. m\left( \epsilon+t\left[ \epsilon,X \right] \right) \right| _{t=0}.
\eqend
 In parallel with the charged particle case, it can be shown that up to coboundaries (this concept will soon be explained), eq. \Ref{eq:ST} is still true in 1 space dimension if $\hat{d\Gamma }(X)$ is replaced with $G(X,\epsilon )$. In fact, it can even be replaced with $G(X,\epsilon _0)$, where the current $\hat{d\Gamma }(X, \epsilon _0)$ is defined as in \Ref{eq:DGA} but with $P$ replaced with the orthogonal projection $P_0$ belonging to the sign of the free hamiltonian $\epsilon _0$. This is equivalent will the well-known fact (at least for charged particles) that the 1-dimensional Schwinger term does not depend on the gauge potential. The difference between $\hat{d\Gamma }(X)$ and $G(X,\epsilon )$ becomes apparent first in higher dimensions. The basic ideas are the same for all odd dimensions $\geq 3$ so we restrict to only consider the case for $d=3$. As in \cite{ME}, the physically important implementer in higher dimensions is given by 
\eq
\label{eq:AAA}
\tilde{G}(X,\epsilon ):=\U ^{\ast }(\epsilon )G(X,\epsilon _0)\U (\epsilon ).
\eqend
 The unitary operator $\U (\epsilon )$ acting on $\F _{\gamma }(h)$ is by definition the operator which maps the vacuum vector in the Fock space determined by $\epsilon $ to the vacuum vector in the Fock space determined by $\epsilon _0$. Thus, $\U (\epsilon )$ is only well defined for $\epsilon $ such that $\epsilon -\epsilon _0\in\B _2$. If $X\in g_2$, then the same is true for $G(X,\epsilon )$ and $\tilde{G}(X,\epsilon )$. To circumvent the  problem of restriction of $X$ and $\epsilon $, we consider the second quantized operators in \Ref{eq:AAA} as sesquilinear forms \cite{R2} and introduce a renormalized commutator as in \cite{ELL}. This in analog with the situation for charged particles and allows a parallel treatment to ref. \cite{ME}. It gives the 3-dimensional Schwinger term:
\eq
\label{eq:ST2}
c(X,Y;\epsilon )=-\frac{1}{16}\mbox{\rm Str}\left( (\epsilon -\epsilon _0)^2\epsilon _0\Big[ \left[ \epsilon _0,X\right] ,\left[ \epsilon _0,Y\right]\Big]\right) .
\eqend
 It can be shown that $\epsilon -\epsilon _0\in\B _4$ for $d=3$ and this, together with the fact that $[\epsilon _0,X]\in\B _4$, implies that the right hand side is well defined. For the case $d=1$ the Schwinger term in eq. \Ref{eq:ST} can be rewritten in terms of $\epsilon _0$ according to: 
\eq
\label{eq:ST8}
c(X,Y;\epsilon _0)=-\frac{1}{4}\mbox{\rm Str}\left( \epsilon _0\left[ \epsilon _0,X\right] \left[ \epsilon _0,Y\right]\right) .
\eqend
From eq. \Ref{eq:ST2} and \Ref{eq:ST8}, local forms of the Schwinger terms can be calculated as in ref. \cite{ME}.

 Finally, we would like to mention the cohomological meaning of the Schwinger term. A coboundary is the change in the Schwinger term if adding a function $\xi =\xi (X,\epsilon )$ to $\tilde{G}(X,\epsilon )$, namely: 
\eq
(\delta \xi )(X,Y)=\xi ([X,Y])-\cL _X\xi (Y)+(-1)^{<X,Y>}\cL _Y\xi (X). 
\eqend
 The defining equation for the Schwinger term implies that it has to satisfy:
\eq
c([X,Y],Z;\epsilon )-\cL _Xc(Y,Z;\epsilon )+\mbox{graded cyclic perm.}=0,
\eqend
the so-called cocycle relation. If restricting to local expressions, then the set of cocycles modular the set of coboundaries defines the physically important cohomology for the Schwinger term. 

\section{Including Grassmann numbers}
\label{sec:IGN}
 In this section we extend the formalism to contain Grassmann numbers. The motivation for this is to make the formalism compatible with supersymmetric quantum mechanics models, as the one considered in \cite{GP}. 
 For notations and basic definitions, see the Appendix.

We use the notation $B_N$ for the associative and commutative $\ast$-
superalgebra over ${\bf C}$ generated by $\{\theta _i\}_{i=1}^N, 
\theta _i=\theta _i^\ast , \mbox{\small{deg}}(\theta _i )=\bar{1}$. 
$B_N$ is $2^N$-dimensional and can be equipped with a homogeneous basis 
$\{\beta _\mu\}_{\mu =1}^{2^N-1}$. 
The (left) $B_N$-module $B_N\otimes h$ is by definition the linear space 
spanned by the elements $f=  \beta _\mu\otimes _{\bf C}f_\mu ,f_\mu\in h $. 
For convenience, the symbol for the tensorproduct will often be omitted. 
 We define the grading according to 
\eq
\mbox{\small{deg}}(\beta _\mu f_\mu )= \mbox{\small{deg}}(\beta _\mu )+ 
\mbox{\small{deg}}(f_\mu )
\eqend
and we extend the inner product in $h$ to a product in $B_N\otimes h$ 
according to 
\eq
(\beta _\mu f_{1,\mu } , \beta _\nu f_{2,\nu })= 
(-1)^{<\beta _\mu + \beta _\nu ,f_{1, \mu } >} \beta _\mu^\ast \beta _\nu 
(f_{1,\mu },f_{2,\nu }).
\eqend
It carries a left action of $B_N$ according to
\eq
\beta _\nu (\beta _\mu f_\mu )= (\beta _\nu\beta _\mu )f_\mu .
\eqend
 and a right action according to
\eq
(\beta _\mu f_\mu )\beta _\nu = (-1)^{<\beta _\nu , \beta _\mu +
f_{1, \mu } >} (\beta _\nu\beta _\mu )f_\mu .
\eqend
Of interest is also the $\ast$-superalgebra $B_N\otimes {\cal A}$, 
for ${\cal A}$ a complex involutive superalgebra. It is the algebra 
generated by the elements $A= \beta _\mu\otimes _{\bf C}A_\mu ,A_\mu
\in {\cal A} $. 
The algebraic product is defined by
\eq
(\beta _\mu\otimes _{\bf C}A_{1,\mu })(\beta _\nu\otimes _{\bf C}A_{2,
\nu })=(-1)^{<\beta _\nu ,A_{1, \mu } >}\beta _\mu \beta _\nu 
\otimes _{\bf C}(A_{1,\mu }A_{2,\nu }),
\eqend
the grading by 
\eq
\mbox{\small{deg}}(\beta _\mu A_\mu )=
\mbox{\small{deg}}(\beta _\mu )+ 
\mbox{\small{deg}}(A_\mu )
\eqend
and the involution by 
\eq
(\beta _\mu A_\mu )^\ast =(-1)^{<\beta _\mu ,A_\mu >}
\beta _\mu^\ast A_\mu^\ast .
\eqend
 Furthermore, in the case of ${\cal A}\subset{\cal A} (h)$, the set of 
linear operators on $h$, we define an action of  $B_N\otimes {\cal A}$ 
on $B_N\otimes h$ according to 
\eq
\label{eq:AAH}
(\beta _\mu\otimes _{\bf C}A_{\mu })\cdot (\beta _\nu \otimes _{\bf C} 
f_{\nu })=(-1)^{<\beta _\nu ,A_\mu >}\beta _\mu \beta _\nu 
\otimes _{\bf C}(
A_{\mu }f_{\nu }).
\eqend 
 This implies
\eq
(f_1,Af_2)=(A^\ast f_1, f_2) .
\eqend
 The trace $\mbox{Tr}_{B_N\otimes{\cal A}}$ on $B_N\otimes{\cal A}
\subset B_N\otimes{\cal A} (h)$ is the 
linear operator defined by 
\eq
\mbox{Tr}_{B_N\otimes{\cal A}}(\beta _\mu A_\mu )=\beta _\mu 
\mbox{Tr}_{\cal A}(A_\mu ).
\eqend
Now, the structure of the formalism developed in the earlier sections is close 
to the one that is obtained by including Grassmann numbers.
 We will go through 
the basic definitions and point out where things differ when Grassmann 
numbers are included. 

$B_N \otimes _{\bf C} 
{\cal A}_{SSR}$ is generated by the operators $B^{\dagger }(\beta _\mu f_{\mu })
\equiv \beta _\mu B^{\dagger }(f_{\mu }),B(\beta _\mu f_{\mu })
$ $\equiv (-1)^{<\beta _\mu ,f_\mu >}\beta _\mu^\ast B(f_{\mu }) $ 
and an identity. The corresponding relations of \Ref{eq:SSR} are fulfilled  
except for the last which is replaced by 
\eq
B(f)  =  B^{\dagger }(qJ Kf) ,
\eqend
where $f$ is a linear combination of elements $\beta _\mu f_{\mu }$ and 
 $K$ is defined by 
\eq
K\beta _\mu f_{\mu }=(-1)^{<\beta _\mu ,f_\mu >}\beta _\mu^\ast f_{\mu }.
\eqend
 Similar, $B_N \otimes _{\bf C} {\cal A}_{CSR}$ 
is generated by the operators $a^{\dagger }(\beta _\mu \tilde{f}_{\mu })
\equiv \beta _\mu a^{\dagger }(\tilde{f}_{\mu })$,  
$a(\beta _\mu \tilde{f}_{\mu })
\equiv (-1)^{<\beta _\mu ,\tilde{f}_\mu >}\beta _\mu^\ast 
a(\tilde{f}_{\mu }) $ 
and an identity. They obey the corresponding relations of \Ref{eq:CSR}. 
 The $\ast$-isomorphism is defined 
as in  \Ref{eq:SEC}, but with $J$ replaced by the non-antiunitary 
operator $JK$.

 Using the representations of ${\cal A}_{CSR}$ defined in section 
\Ref{sec:SSR} we may follow the recipe of \Ref{eq:AAH} to define a 
corresponding representation 
of $B_N \otimes _{\bf C} {\cal A}_{CSR}$ in $B_N\otimes \F _{\gamma }(\tilde{h})$. 
As before, this provides us with a representation of 
$B_N \otimes _{\bf C} {\cal A}_{SSR}$. The defining equation is  
\Ref{eq:SE0} with $J$ replaced by $JK$.
 It is now straight forward to check that everything said in 
section  \Ref{sec:IAS} goes through provided we replace every space 
with $B_N$ tensored with it and make the substitution $J\longrightarrow JK$.
 The results of section \Ref{sec:FBT} can be extended in a similar way, see \cite{GL2} for the corresponding case for charged particles.

\section*{Acknowledgements}
 I thank Edwin Langmann for allowing me to use unpublished material from \cite{EL} on which this paper is based.

 \section*{Appendix: Super algebras and super vector \\spaces}
\label{sec:PSDO}
Some basic facts about  
${\bf{Z}}_2$-graded vector spaces and algebras will be summarized here. 
An element $v$ in a  
${\bf{Z}}_2$-graded vector space $V=V_{\bar{0}}\oplus V_{\bar{1}}$ is said 
to be homogeneous of  degree  $\alpha $, $\mbox{\small{deg}}(v)
=\alpha$, if $v\in V_{\alpha }$,  $\alpha \in  {\bf{Z}}_2\equiv 
\{ \bar{0}, \bar{1} \} $. 
If $V$ is also an 
algebra with grading preserving multiplication, i.e. 
 $v\in V_{\alpha }, w\in V_{\beta } \Rightarrow vw\in V_{\alpha + \beta}$
, then it is called a  
${\bf{Z}}_2$-graded algebra.
We define the supercommutator 
$\left[ \cdot , \cdot \right] _s :$ $V\times V \rightarrow V$ by
\eq
\label{eq:SC}
\left[ v , w\right] _s = v w -(-1)^{<v,w>}wv.
\eqend
where the notation $<v,w>$ means $\mbox{\small{deg}}(v)\cdot
\mbox{\small{deg}}(w)$.
 This definition is by linearity 
also well defined for non-homogeneous elements.
 Equipped with the supercommutator, $V$ becomes a Lie superalgebra.
Every linear operator $X$ on $V$ can be written 
in matrix form 
\eq
\label{eq:MX}
X=\left(
 \begin{array}{cc}
 X_{\bar{0}\bar{0}} &  X_{\bar{0}\bar{1}} \nonu
  X_{\bar{1}\bar{0}} &  X_{\bar{1}\bar{1}} 
 \end{array} \right)
\eqend
corresponding to the decomposition $V=V_{\bar{0}}\oplus V_{\bar{1}}$. 
Then  $\mbox{\small{deg}}( X_{\alpha\beta })=\alpha +\beta$ defines a 
grading which provides every algebra of linear operators on $V$ with a
${\bf{Z}}_2$-structure.

\end{document}